\documentclass[12pt]{article}

\usepackage[a4paper]{geometry}

\usepackage{setspace}
 
\usepackage{mathptmx}
\usepackage{newtxtext,newtxmath}

\usepackage{amsmath}

\usepackage{bm}

\usepackage{graphics}
\usepackage{epsfig}
\usepackage{colortbl}
\usepackage{color}
\usepackage{nicefrac}
\usepackage{mathrsfs}
\usepackage{bbm}

\usepackage{subfig}
\usepackage{booktabs}

\usepackage[breaklinks=true,colorlinks=true,linkcolor=blue,urlcolor=blue,citecolor=blue]{hyperref}

\RequirePackage{color}
\usepackage{soul}

\begin{document}

\begin{titlepage}
	
	\vskip 1.5 cm
	
		\begin{center}
	\Large On the new universality class in structurally disordered $n$-vector model with long-range interactions
	\end{center}
	
	\vskip 2.0 cm
	\centerline{{\bf{Dmytro Shapoval},}  $^{1, 2}$ 
		{\bf{Maxym Dudka},} $^{1, 2}$
		{\bf{Yurij Holovatch}} $^{1, 2,3}$
		}	\vskip 0.5 cm
	\begin{center}	
		$^1$
		Institute of Condensed Matter Physics, National Academy of Sciences of Ukraine, UA -- 79011 Lviv, Ukraine
	\\ \vspace{0.5cm}
		$^2$ 
		${\mathbb L}^4$ Collaboration \& Doctoral College for the Statistical Physics of Complex Systems, Leipzig-Lorraine-Lviv-Coventry, Europe
		\\ \vspace{0.5cm}
		$^3$ 
		Centre for Fluid and Complex Systems, Coventry University, Coventry, CV1 5FB, United Kingdom

	\end{center}

	\begin{abstract}
		We study a stability border of a region where nontrivial critical behaviour of an $n$-vector model with long-range power-law decaying interactions is 
		induced by the presence of a structural disorder (e.g. weak quenched dilution). This border is given by the marginal dimension of the order parameter $n_c$ dependent on space dimension, $d$, and a control parameter of {the} interaction decay, $\sigma$, below which the model belongs to the new dilution-induced universality class. Exploiting the Harris criterion and recent field-theoretical renormalization group results for the pure model with long-range interactions we get $n_c$ as a three loop $\epsilon=2\sigma-d$-expansion. We provide numerical values for $n_c$ applying series resummation methods. Our results show that not only the Ising systems ($n=1$) can belong to the new disorder-induced long-range universality class at $d=2$ and $d=3$.
			
	\end{abstract}
\end{titlepage}

\clearpage

\section{Introduction}

Year 2022 marks the 110th birth anniversary of Oleksandr (a.k.a. A.S.) Davydov, an outstanding physicist, known for his
seminal contributions in the fields of solid state theory, nuclear physics and biophysics, for
many years he served as a director of the Bogolyubov Institute for Theoretical Physics in Kyiv. The authors of this paper 
who studied physics also from O. Davydov's books \cite{Davydov13,Davydov71,Davydov85} consider as a great honour
to contribute to the Festschrift prepared on this occasion. In our paper we use the perturbative field theoretical 
renormalization approach refined by the resummation of asymptotic series expansions to study universal features of
criticality. Although such problems were beyond the focus of attention of O. Davydov, 
the concepts called for their analysis: Symmetry, Space dimension, Range of interaction belong to the central ones
in physics.
In the paper we show how their interplay defines universal features of one of the key models currently used
to  understand quantitatively and to describe qualitatively the critical behaviour  in condensed matter and beyond. 
Therefore, conceptually the results presented in this paper are related to those discussed in O. Davydov's
seminal works.  For this reason we have chosen to present these results here.

Since inter-particle forces in various physical, chemical, and biological systems are often of a long-range nature, models with 
long-range interaction attract much attention. They have found their applications in studies of gravitational, dipolar, cold Coulomb systems, problems in plasma, atomic and nuclear physics,  hydrodynamics and geophysical fluid mechanics (see \cite{review,lectures0,lectures} and reference therein). Systems with long-range interactions possess properties that differ from those with short-range interactions. To give an example, even weak long-range interactions effectively modify the critical properties  and may induce the long-range order in one-dimensional systems \cite{lectures0,lectures}.

In this paper we will discuss possible changes in  the critical behaviour of a many-particle system  
caused by mutual effects of long-range interactions and structural disorder. To this end, we 
will consider the -- now standard -- $n$-vector spin model with the Hamiltonian:
\begin{equation}\label{ham_spin}
{\cal H} =  - \frac{1}{2}\sum_{{\bf x},{\bf x}'} J(|{\bf x}-{\bf x}'|)
\vec{S}_{\bf x}\vec{S}_{{\bf x}'}\, 
\end{equation}
that describes a system of classical $n$-component vectors ('spins') 
$\vec{S}_{\bf x}= (S^1_{\bf x}, S^2_{\bf x},\dots, S^n_{\bf x}, )$ located at sites ${\bf x}$ of a $d$-dimensional 
lattice and interacting via the distance-dependent potential $J(x)$. The r.h.s of Eq. (\ref{ham_spin}) contains
a scalar product of spins and the sums over ${\bf x}, {\bf x'}$ span all 
lattice sites. Influence of long-range interactions on 
the critical  behavior is usually  exemplified by the
power-law decaying  interaction: 
\begin{equation}\label{int}
J(x)\sim x^{-d-\sigma},
\end{equation} 
where $\sigma > 0$ is a control parameter of the interaction decay.

As we discuss in more details below, in the case of a regular (non-disordered) lattices,
the critical behaviour of the model (\ref{ham_spin}) is governed by the triple of parameters 
$(d, n, \sigma)$: depending on their values, the model may manifest a low-temperature
long-range order that emerges as a second-order phase transition. 
As long as the model Hamiltonian (\ref{ham_spin}) is formulated in terms of elementary magnets -- `spins',
the long-range ordered phase is usually associated with an emergence of spontaneous magnetization.
We will use such magnetic terminology too, however let us note that the model itself as well
as our discussion 
without the loss of generality concern much more wide range of types of ordering \cite{rgbooks,Order_disorder}
in physics and beyond.
The transition to the ordered `magnetic' phase is
characterised by certain universal (i.e. independent on specific system details) 
features. It is said, that it belongs to certain {\em universality class}. Systems,
that belong to the same universality class share the values of critical exponents,
amplitude ratios, scaling functions. Our goal in this paper is to show how these
universal features are changed if instead of a regular lattice structure, one 
considers the disordered one. Disorder in the lattice structure may be imposed e.g. by
dilution, when a part of the lattice sites in (\ref{ham_spin}) are not
occupied by spins. Such situation mimics randomness and non-regularities that
are so often met in nature and attract much interest in modern theory of critical
phenomena (see e.g. \cite{Order_disorder} and references therein.) To quantify
our analysis, we will calculate the {\em marginal dimension}
$n_c(\sigma)$: for given space dimension $d$ it discriminates between different
universality classes. The rest of the paper is organized as follows. In the next Section
\label{Review} we give a short review of the results present so far, in
Section  \ref{field_theory} we describe the field-theoretical renormalization group 
picture of the critical behaviour for the long-range interacting $n$-vector model with 
disorder. The results for $n_c$ are given in Section \ref{marginal} and we summarize our 
study in the last Section \ref{conclusions}. Some lengthy expressions are given in the Appendix.

\section{Review}
\label{review}

In this section we briefly review some results relevant for our
analysis. To proceed further, we explain terminology, used 
throughout the paper. The $n$-vector model (\ref{ham_spin})
with short-range interactions
-- since the corresponding free energy is invariant with respect 
to rotations in the $n$-dimensional magnetization space it is also called the 
$O(n)$-symmetric model -- 
manifests the second order phase transition for the lattice space dimension
$d>d_{lc}$. The lower critical dimension $d_{lc}=1$ for the discrete
(Ising) case $n=1$ whereas $d_{lc}=2$ for $n>1$. Critical exponents and other
universal properties of the short-range $n$-vector model depend on $n$ and $d$ 
in a non-trivial way in the region $d_{lc}<d\leq d_{uc}$. It is said that they
belong to the short-range universality class. For $d$ larger
than the upper critical dimension $d_{uc}=4$ the model is governed by the
mean-field exponents, see \cite{Berche22} for more details. Introducing the long-range interaction (\ref{int}) drastically changes
the picture of the critical behaviour of the $n$-vector model (\ref{ham_spin}). Calculations 
performed for the three-dimensional spherical model \cite{Joyce1966} 
(it corresponds to the $n$-vector model at $d=3,n=\infty$) show that for $\sigma>2$ the critical properties 
are governed by the short-range critical exponents, while for  $\sigma<2$ one has two regimes 
depending on the value of $\sigma$: with mean-field critical exponents and with the 
$\sigma$-dependent ones. One-dimensional Ising model ($d=1,n=1$) with interaction (\ref{int}) 
was proven to have phase transition to the long-range-ordered  phase at non-zero temperature \cite{Dyson1969}. 
Field-theoretical renormalization group (RG) analysis of the long-range 
interacting $n$-vector model  
  gives three universality classes in dependence 
on $\sigma$ \cite{Fisher1972}: (i) the mean-field critical behavior   for $\sigma\le d/2$, (ii) {\em the
short-range universality class} for $\sigma\ge 2$ with critical exponents coinciding with 
those of the model with short-range interactions, (iii) {\em the long-range universality class} 
for $d/2<\sigma< 2$, where critical exponents depend on $\sigma$. Later it was established 
that the actual border between the short-range and the long-range universality classes lies at  
$\sigma=2-\eta_{SR}$ \cite{Sak1973,Sak1977} rather than at $\sigma=2$, $\eta_{SR}$ is the
pair correlation function
critical exponent of the short-range model.
Such picture  was corroborated by other approaches including non-perturbative variant of RG 
(NPRG) \cite{Defenu2015}, Monte Carlo simulations \cite{Luijten2002} and conformal bootstrap 
\cite{Behan2019} (for other references and discussion see the review \cite{Defenu2020}).

Another factor discussed in this paper is the structural disorder and its impact on the critical behaviour.
The influence of structural inhomogeneity on the universal properties of physical systems continues to be a 
hot research topic both for academic and practical reasons, since almost all materials are characterized 
by a certain degree of disorder in their structure. Structural inhomogeneities  in magnetic systems 
are of different nature, which in turn may lead to 
different changes in  critical behavior. In the case of strong structural disorder, randomness is accompanied by frustration
and percolation
 effects and often leads to absence of the long-range magnetic order. The case of weak structural disorder is not that obvious. Here we focus specifically on presence of the weak quenched disorder in 
lattice structure, which may be implemented into model (\ref{ham_spin}) via dilution by point-like uncorrelated 
(or short-range correlated) quenched non-magnetic inhmogeneities. 
Its relevance for the critical behaviour  is given by the Harris criterion \cite{Harris1974}.
The criterion states that the structural disorder (quenched dilution) leads to a new universality class of the magnetic phase transition only if the heat capacity critical exponent of the undiluted  (pure)   system
is positive, $\alpha_{p} > 0$, i.e. if the heat capacity of the pure system diverges at the critical point. Correspondingly, the disorder is irrelevant if  $\alpha_{p} < 0$.
 For the short-range  $n$-vector model at $d=3$
$\alpha_{p} > 0$ for  $n=1$ (Ising model), whereas $\alpha_{p} < 0$ for  $n\geq2$, therefore due to the
Harris criterion the diluted $n$-vector model at $n\ge 2$ shares the universal properties of 
its undiluted counterpart \cite{Pelissetto02,Holovatch02}. Unlike the short-range $n$-vector model, 
the  long-range one manifests the new universality class induced by dilution also in the region $n\ge 2$
as it was shown in the RG study of Ref.  \cite{Yamazaki1978} within two-loop approximation. This result was also corroborated  by
the low-temperature RG \cite{Li1981}. However, the estimates of the regions of values $(d, n,\sigma)$, where the new disordered long-range
universality class governs the critical behaviour were not satisfactory \cite{Yamazaki1978}. 
The perturbative RG expansions having zero radius of convergence, additional resummation procedures
have to be applied in order to get reliable numerical data on their basis \cite{rgbooks}.
Especially it concerns the Ising model ($n=1$), where degeneracy of the  RG equations (similarly as in the short-range case \cite{Pelissetto02, Holovatch02}) makes 
the expansion parameter to be  $\sqrt{\epsilon}$ ~\cite{Theumann1981}. 
Reliable results for the last case were obtained within a massive renormalization scheme with resummation of the RG functions \cite{Belim2003d}.

A remarkable feature of the Harris criterion is that it allows to forecast structural-disorder-induced changes 
in the universality class of a pure system without explicit  calculation of the RG functions for the diluted one.
Indeed, if the structural disorder changes the universality class only when the heat capacity of the pure
system diverges (i.e. when
$\alpha_{p} > 0)$, one can use the  condition $\alpha_{p} = 0$ as an equation to define parameters ($d,\sigma,n$)
that discriminate between different universality classes. For given space dimension $d$ such equation defines 
the so-called marginal order-parameter dimension $n_c(\sigma)$. Similar to critical exponents and critical amplitude 
ratios, the marginal dimensions are universal quantities, reachable in experiments and numerical simulations
and are the subject of intensive studies \cite{Bervillier86,Dudka01,cubic,mn,frustrated,Dudka2012}.
In the next sections we will calculate  the marginal dimension $n_c(\sigma)$ for the diluted long-range $n$-vector  model 
at space dimensions $d=2$ and $d=3$. This marginal dimension line 
in the $(n,\,\sigma)$ parametric plane
defines a border of stability  between the pure long-range 
and the disordered long-range universality classes.  To this end we will make use of the 
recent three-loop RG  results for the critical exponents of the pure $n$-vector model with long-range 
power-law decaying interactions \cite{Benedetti2020}.

\section{Field-theoretical  renormalization group description}
\label{field_theory}

The progress achieved in quantitative understanding and qualitative description of critical phenomena 
to a large extend is due to application of  RG methods \cite{rgbooks}. In the field-theoretical
RG approach, the critical properties of the $n$-vector model (\ref{ham_spin}) with the power-law decaying long-range 
interactions (\ref{int})  are described by analysing the effective  Hamiltonian \cite{Suzuki1972}: 
    \begin{equation}
        \label{ham0}
            {\mathcal H}=\int d^d x \left\{\frac{1}{2}\left((\nabla^{\sigma/2}{\vec\varphi})^2+r_0{\vec\varphi}^2\right)+\frac{u_0}{4!}({\vec \varphi}^2)^2\right\},
    \end{equation}
where $\vec{\varphi}=\vec{\varphi}({\bf x})=\{\varphi_1({\bf x}),\dots,\varphi_n({\bf x})\}$ is an $n$-component field,  $u_0$ is unrenormalized coupling, $r_0$ defines the temperature distance to the critical point, and $\nabla^{\sigma/2}$ is a symbolic notation for the fractional derivative. The last is defined via  its action in the momentum 
space and leads to the propagator term   $q^{\sigma}$ rather than  $q^2$ as in the case of short-range interactions. Power counting  gives the upper critical dimension $d_{uc}=2\sigma$, which at $\sigma=2$ coincides with the traditional $d_{uc}=4$. The effective Hamiltonian (\ref{ham0})  is relevant for the case  $0<\sigma<2-\eta_{SR}$.  To study crossover to the short-range case $\sigma\to 2$ one should  include the traditional term $(\nabla{\vec\varphi})^2$ into (\ref{ham0}).     

 In the field-theoretical RG approach, a critical point corresponds to a reachable and stable fixed point (FP) of the RG transformation.
 It has been found \cite{Fisher1972} that the nontrivial FP determining new long-range universality class is stable  for $d/2<\sigma$. Critical exponents within this universality class were calculated
 in $\epsilon=2\sigma-d$-expansion up to order $\epsilon^2$ \cite{Fisher1972,Suzuki1972,Yamazaki1977}  and up to order $1/n$ \cite{suzuki} in $1/n$-expansion. Estimates for the critical exponents of three-dimensional systems were obtained within the massive renormalization approach completed by resummation in two-loop approximation \cite{Belim2003}. The  RG results in three-loop approximation were obtained only recently \cite{Benedetti2020}.  Monte Carlo estimates for the critical exponents in this universality class have been obtained 
 only for the $n=1$ Ising case mainly in one and two space dimensions (for collection of references see \cite{Benedetti2020}).  Only a few Monte Carlo results are available for the three-dimensional case \cite{Belim2016}. 

The presence of uncorrelated non–magnetic impurities (weak quenched structural disorder) is usually modeled by
fluctuations of the local phase transition temperature \cite{GrinsteinLuther}. Introducing $\phi=\phi({\bf x})$ as the field of local 
critical temperature fluctuations, one obtains the following effective Hamiltonian for the structurally disordered system:
    \begin{equation}
        \label{ham}
	        {\mathcal H}=\int d^d x \left\{\frac{1}{2}\left((\nabla^{\sigma/2}{\vec\varphi})^2+\left(r_0+\phi\right){\vec\varphi}^2\right)+\frac{u_0}{4!}({\vec \varphi}^2)^2\right\}\, ,
    \end{equation}
where the random variable $\phi$  has a Gaussian distribution with zero mean and a correlator containing 
the second coupling $v_0$:    
    \begin{equation}
        \label{average}
            \left<\phi({\bf x})\right>=0,\qquad\left<\phi({\bf x})\phi({\bf x'})\right>=v_0\delta({\bf x}-\bf{ x'})\, .
    \end{equation}
The angular brackets $\langle\dots\rangle$ indicate an average over   
the random variable $\phi$ distribution.

The RG picture  for the disordered long-range model (\ref{ham})   is similar to {that
for} its short-range analog  \cite{Pelissetto02,Holovatch02}. In the parametric space of couplings  $(u,\,v)$ the critical properties of the model  (\ref{ham}) are governed by four FPs $(u^*,\,v^*)$  in dependence on values $\epsilon=2\sigma-d$ and $n$~\cite{Yamazaki1978, Theumann1981}: Gaussian FP $(u^*=0,\,v^*=0)$, unphysical FP $(u^*=0,\,v^*\not=0)$, Heisenberg long-range  or pure long-range FP $(u^*=0,\,v^*\not=0)$  and disordered long-range FP $(u^*\not=0,\,v^*\not=0)$. The Gaussian FP is always unstable below critical $d_u=2\sigma$ ($\epsilon>0$) while the unphysical FP is always stable in this case, however, it is not accessible from initial conditions appropriate for the model described by (\ref{ham}) and (\ref{average}).  For $\epsilon>0$ and $ n > n_c$, the long-range Heisenberg FP is stable and the disordered one is unstable,  while for $n < n_c$ 
the FPs swap their stability.  Therefore for $n > n_c$ the universal critical exponents of the diluted model  (\ref{ham}) coincide with those of the model  (\ref{ham0}). For $n < n_c$  model  (\ref{ham}) belongs to the new disorder-induced long-range universality class.  The border between these two regimes is determined by the marginal dimension {$n_c(d,\sigma)$.} 

So far, the value of $n_c$ for the long-range $n$-vector model was known within the two-loop approximation~\cite{Chen2003}, with the result
    \begin{equation}
        \label{2loop}
        	n_c=4-4[\psi(1) - 2\psi(\sigma/2) + \psi(\sigma)]\epsilon + O(\epsilon^2)\, ,
    \end{equation}
where $\psi(x)$ is the digamma function. The asymptotic nature of this series together with its shortness made it difficult to get reliable numerical estimates 
on its basis.
In the next section we will proceed getting the  next order of the $\epsilon$-expansion and delivering numerical  estimates for $n_c(\sigma)$ at certain space dimensions with the help of resummation procedures.

\section{Calculation of the marginal dimension}
\label{marginal}

As it has been mentioned above, the marginal dimension $n_c$  of a weakly diluted $n$-vector model with power-law decaying interactions can be obtained on the
base of the critical exponents for the undiluted model. As a consequence of the Harris criterion, the master equation for determining $n_c$ reads: 
    \begin{equation}
        \label{alp}
        {\alpha_p}(n_c,d,\sigma) = 0. 
    \end{equation}
The treatment of Eq.~(\ref{alp}) by means of the field-theoretical RG approach can be performed in various schemes. Here we will exploit the results of dimensional regularisation with the minimal subtraction \cite{tHooft} allowing to obtain quantities of interest by familiar $\epsilon$-expansion with $\epsilon = 2\sigma -d$ in our case. To get $n_c(d,\sigma) $ we  use the hyperscaling relation ${\alpha_p}=2-d{\nu_p}$  and  $\epsilon$-expansion for the critical exponent {$\nu_p$} of the $n$-vector model with long-range interactions, which is known in the three-loop  approximation  \cite{Benedetti2020} in the following form: 
    \begin{align} 
		{\nu_p}^{-1} &= \,
		 \sigma \, - \, \frac{(n+2)}{n+8} \, \epsilon \, + \, \frac{(n+2)(7n+20)\alpha_{S,0}}{(n+8)^3} \, \epsilon^2 + \, \frac{(n+2)\epsilon^3}{(n+8)^5} \bigg[ -4(5n+22)(7n+20) \alpha_{S,0}^2\nonumber\\
		&\quad  + (n+8)^2(7n+20) (\alpha_{S,1} -2 \alpha_{D,1} \alpha_{S,0}) + (n+8) \Big({-}8(n-1) \alpha_{T} + 2(n^2+20n+60) \alpha_U \nonumber\\
		&\quad  + 2(n^2+24n+56) \alpha_{I_1}
		+ (5n^2+28n+48) \alpha_{I_2} + (5n +22) \alpha_{I_4} \Big) \bigg] \, + \, {O}(\epsilon^4)
		 \, , 
	\end{align}
where $\alpha_K$ with $K=\{S,D,I_1,I_2,I_4,T,U\}$ are expressed in terms of the   loop integrals (for details see \cite{Benedetti2020}). $\alpha_{S,i}$ and $\alpha_{D,i}$ are the coefficients at $\epsilon^i$ in the $\epsilon$-expansion series of  $\alpha_S$ and $\alpha_D$.  Explicit expressions for  $\alpha_K$ are given in the Appendix. We get the following expression for  $n_c$:
    \begin{equation}
        \label{nc}
            n_c(d,\sigma)=4 + 4 \alpha_{S,0}\epsilon +  (56 \alpha_{I_1} + 40\alpha_{I_2} + 7 \alpha_{I_4} - 192 \alpha_{D,1} \alpha_{S,0} - 56 \alpha_{S,0}^2 + 96 \alpha_{S,1} - 4 \alpha_T + 52 \alpha_U) \frac{\epsilon^2}{24}+O(\epsilon^3)\, .
    \end{equation}
Taking into account the explicit expression for $\alpha_{S,0} $ one can check that up to the first order in $\epsilon$ {Eq. (\ref{nc})} 
coincides with the two-loop result (\ref{2loop}). 
\begin{figure}[!htbp]
	\begin{center}
		\includegraphics[width=.38\paperwidth]{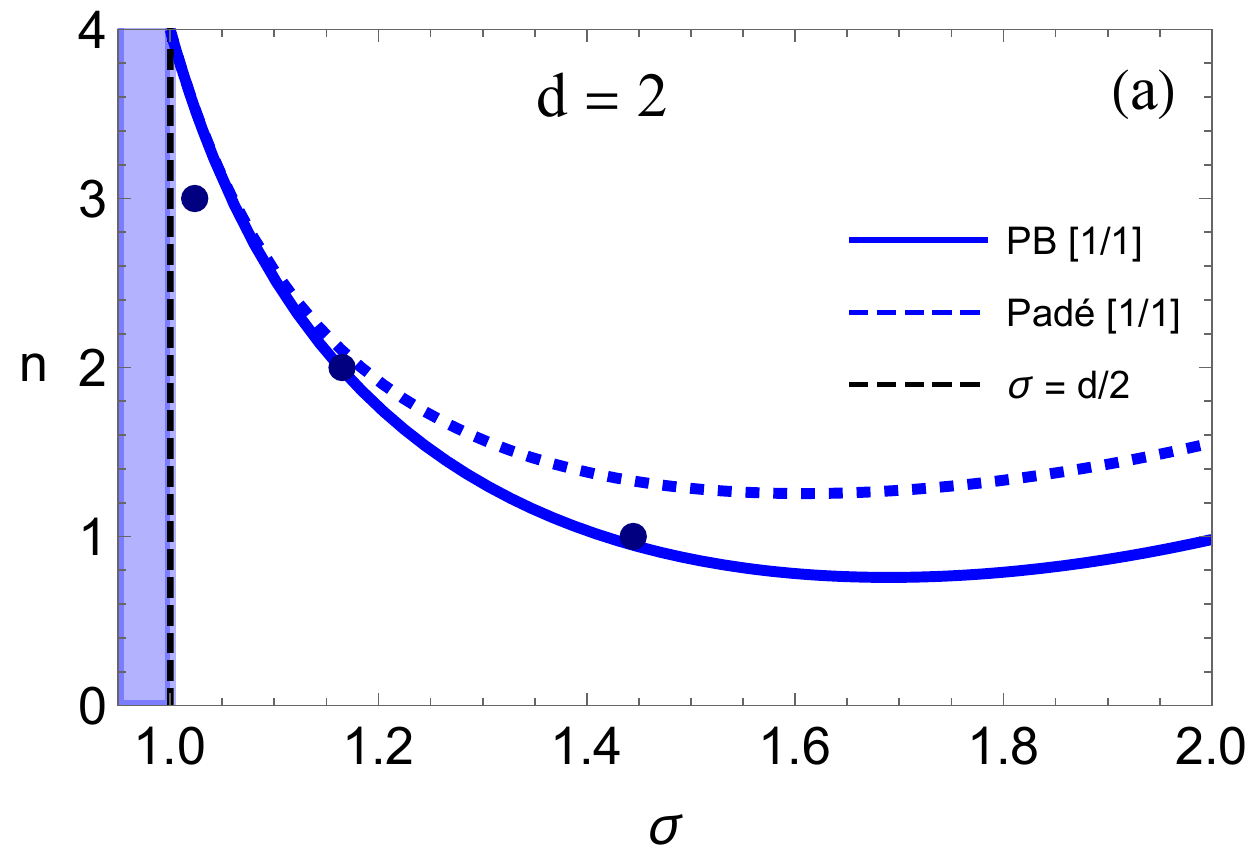}
		\includegraphics[width=.38\paperwidth]{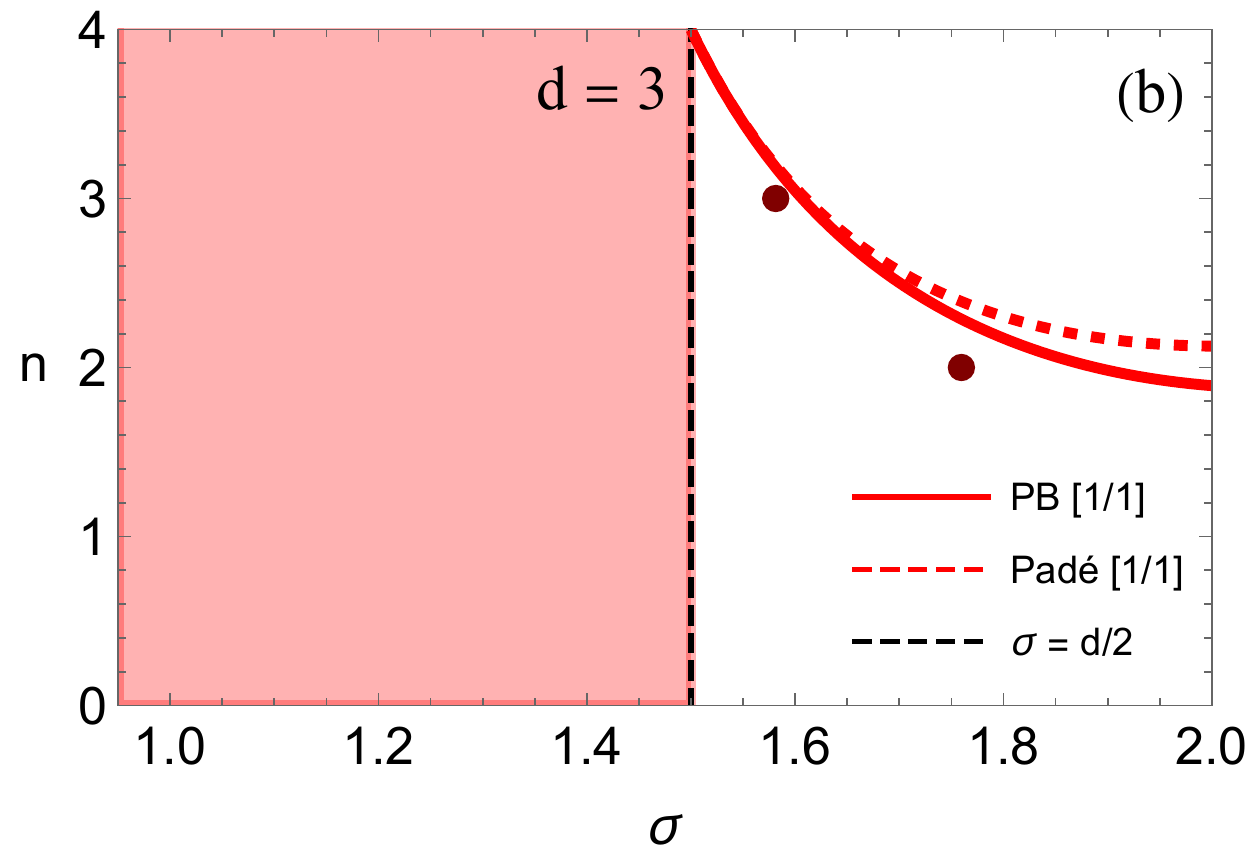}
	\end{center}
	\caption{Resummed values of $n_c(\sigma)$ obtained by the Pad\'e-approximant [1/1] (dashed lines) and  Pad\'e-Borel resummation (solid lines)  for $d=2$ (a) and $d=3$ (b). For the region of values $n$ and $\sigma$ above the lines, the pure long-range universality class holds, while the new disorder long-range universality class is induced for the values below the lines. Dots show results that follow from the interpolation of the NPRG data of Ref. \protect\cite{Defenu2015}. 
	The mean-field behaviour holds for $\sigma<d/2$ (regions separated by vertical dashed lines, coloured online).\label{nccc}}
\end{figure}

Formally, the numerical value of $n_c$  at fixed $d$ and $\sigma$ can be calculated from the expansion 
(\ref{nc}) by using  expressions for $\alpha_K$ from the Appendix, recalling that $\epsilon=2\sigma-d$ and substituting the
values of $d$ and $\sigma$. 
However the $\epsilon$-expansions  are known to be asymptotic at best \cite{rgbooks}. 
Therefore one has to apply special resummation procedures to restore their convergence in order to get reliable numerical estimates on their basis. Doing so, we start our analysis by representing series (\ref{nc}) by  means of the diagonal Pad\'e approximant  $[1/1] (\epsilon)$ \cite{Pade}. 
The result for $n_c$  as a function of $\sigma$ is shown by dashed lines in Fig.~\ref{nccc} a,b for fixed $d=2$ and $d=3$,
correspondingly. 
As it was noted in the previous Section,
in the region of $n$ and $\sigma$ above the lines the critical behaviour of the diluted model is the same as for an undiluted model with long-range interactions. For the values of $n$ and $\sigma$  below the lines, the new disorder long-range universality class is induced. It is known that in the short-range case the Pad\'e approximants of the three-loop expansion give values of $n_c$ that exceed the most accurate estimates  \cite{Dudka01}. It seems to be also in the long-range case, since the data of the NPRG approach {(shown by black dots in Fig.~\ref{nccc})} are located below the lines calculated via  Pad\'e approximant.\footnote{These data were obtained  from interpolation of  the NPRG results for critical  exponents of the long-range $n$-vector model \cite{Defenu2015}.} To enhance our estimates we use also a more elaborated Pad\'e-Borel resummation technique \cite{PadeBorel}. 
First, to weaken the factorial growth of the expansion coefficients we  write  the Borel transform for (\ref{nc}) as:
    \begin{equation}  \label{B}
        \sum_{k=0}^{2}n_k \epsilon^k \rightarrow \sum_{k=0}^{2}\frac{n_k \epsilon^k }{k!}\, .
    \end{equation}
An analytical continuation of the Borel transform is achieved by representing it in the form of the diagonal Pad\'e approximant $[1/1]_B (\epsilon)$,
where subscript $B$ is used to distinguish from the Pad\'e approximant of the original series (\ref{nc}). Finally, the resummed function is obtained 
via an inverse Borel transform:
    \begin{equation}  
        n_c^{res}(\epsilon)=\int_0^{\infty} d t {\rm e}^{-t} [1/1]_B (\epsilon t)\, .
    \end{equation}
Results that follow from the  Pad\'e-Borel  resummation are presented in Fig.~\ref{nccc} by solid lines for $d=2$ and $d=3$. For $d=3$ the Pad\'e-Borel  resummation leads to lower values of $n_c(\sigma)$ as compared to those obtained
form the $[1/1]$ Pad\'e approximant. This is a right tendency, as is seen also from comparing our results to the NPRG data, shown by
dots in Fig. \ref{nccc}. As usual with the perturbative expansions, the accuracy of the results  decreases with an increase of
the expansion parameter, in our case it is $\epsilon=2\sigma -d$.
Therefore, our results are less accurate for $d=2$, where the expansion parameter changes within $0\leq \epsilon \leq  2$
for  $1 \leq \sigma \leq  2$. But even then the results of Pad\'e-Borel resummation may serve as reliable estimates up to the moderate values
of $\sigma\simeq1.5$.
A remarkable feature of the plots presented in Figs. \ref{nccc} a,b is that for a certain range of parameters $\sigma$ there are regions in the
$n,\sigma$ plane that correspond to integer values of $n=1,2,3$ and lie below the $n_c(\sigma)$ curve. This means that the new disorder long-range universality class is induced in the $n$-vector model not only for the Ising ($n=1$) but also for the XY ($n=2$) and classical Heisenberg
($n=3$) cases.

\section{Conclusions}
\label{conclusions}

In this study we were interested in the question how the critical behaviour
of a many-particle system is changed under the competing influence of two factors:
type of interaction and structural disorder? To this end, we have considered
the archetype model to describe criticality, an $n$-vector spin model (\ref{ham_spin}), 
and analysed changes in its critical behaviour provided the interaction
between spins is of a long-range nature (\ref{int}) and an underlying lattice
structure is disordered. To be more specific, we considered the case when the weak
quenched structural disorder leads to fluctuations in the local transition
temperature (\ref{average}). The literature available so far \cite{Yamazaki1978,Li1981} reported that
second order phase transition in such a model can belong to the new, disorder induced
long-range universality class. However, such a qualitative conclusion has to be 
supported by quantitative estimates of the region of model parameters where the
new universality class can manifest.

To do so, we have calculated the marginal dimension
$n_c(\sigma)$ of the structurally-disordered long-range interacting $n$-vector model. For given space dimension
$d$ and interaction decay $\sigma$, the model with the order parameter component number $n<n_c$ 
belongs to the new disorder-induced long-range universality class. Based on the recent results
for the critical exponents of the pure long-range $n$-vector model \cite{Benedetti2020}, we used the
Harris criterion to calculate $n_c(\sigma)$ with the record three-loop accuracy, Eq. (\ref{nc}).
This enabled us to apply familiar resummation techniques to evaluate numerical values of the
marginal dimension, as shown in Figs.\ref{nccc} a,b for space dimensions $d=2$ and $d=3$.
Obtained results serve as a solid argument that not only the Ising-like ($n=1$) systems,
but also systems that are described by the XY ($n=2$) and Heisenberg ($n=3$) models belong
to the new universality class for the moderate values of $\sigma$ at space dimesions
$d=2$ and $d=3$.

\section*{Acknowledgements}

We thank the Editors, Yurij Naidyuk, Larissa Brizhik, and Oleksandr Kovalev, for their invitation to contribute to the Festschrift in memory of Oleksandr Davydov. This work was supported in part by the grant of the National Academy of Sciences of Ukraine for research laboratories/groups of young scientists No 07/01-2022(4)) (D.S.).
M.D. thanks members of LPTMC  for their hospitality during  his academic visit in the Sorbonne University, where  the part of this  work was done.

\section*{Appendix}

In this appendix, we list expressions for $\alpha_K$ as they were given in Ref.~\cite{Benedetti2020}
\begin{eqnarray}
	\alpha_{D} \, &=& \, 1 +\frac{\epsilon}{2}\big[\psi(1)-\psi(\tfrac{d}{2}) \big]+\frac{\epsilon^2}{8}\left[\left(\psi(1)-\psi(\tfrac{d}{2})\right)^2+ \psi_1(1)-\psi_1(\tfrac{d}{2})\right] \,, \crcr
	\alpha_{S} \,& = &\,  2\psi( \tfrac{d}{4} ) - \psi( \tfrac{d}{2})-\psi(1)   +\frac{\epsilon}{4}\Big[\left[2\psi(\tfrac{d}{4})-\psi(\tfrac{d}{2})-\psi(1)\right]
	\left[3\psi(1)-5\psi(\tfrac{d}{2})+2\psi(\tfrac{d}{4})\right]   \crcr
	& & + 3\psi_1(1) + 4\psi_1(\tfrac{d}{4})-7\psi_1(\tfrac{d}{2})  -4 J_0(\tfrac{d}{4}) \Big] \, , \crcr
	 \alpha_{U} &=& \alpha_{I_2} =\,  - \psi_1(1)-\psi_1(\tfrac{d}{4})+2\psi_1( \tfrac{d}{2})
	+ J_0(\tfrac{d}{4})\, , \crcr
	\alpha_{T} \, &= &\, \frac{1}{2}\Big[2\psi(\tfrac{d}{4}) - \psi(\tfrac{d}{2})-\psi(1) \Big]^2 + \frac{1}{2}  \psi_1(1)+ \psi_1(\tfrac{d}{4}) - \frac{3}{2} \psi_1(\tfrac{d}{2}) - \, J_0(\tfrac{d}{4}) \, , \crcr
	 \alpha_{I_1} \, &=& \, \frac{3}{2}\left[2\psi( \tfrac{d}{4} ) - \psi( \tfrac{d}{2})-\psi(1)\right]^2
	+ \frac{1}{2} \psi_1(1) 
	-\frac{1}{2}\psi_1(\tfrac{d}{2})\, ,
	\crcr
	 \alpha_{I_4} \, &=& 6 \,\frac{\, \Gamma(1 + \tfrac{d}{4})^3\Gamma(- \tfrac{d}{4})}{
		\, \Gamma(\tfrac{d}{2} )} \;  \Big[  \psi_1(1) -  \psi_1(\tfrac{d}{4})  \Big]\, .  
	\label{eq:alphas}
\end{eqnarray}
In the above expressions  $\psi_i$ are the polygamma functions of order $i$, while $J_0$ is the following sum:
\begin{equation}
	J_0(\tfrac{d}{4})=\frac{1}{\Gamma(\tfrac{d}{4})^2}\sum_{n \geq 1}\frac{\Gamma(n+\tfrac{d}{2})\Gamma(n+ \tfrac{d}{4})^2}{n(n!)\Gamma(\tfrac{d}{2}+2n)}\Big[2\psi(n+1)-\psi(n)-2\psi(n+\tfrac{d}{4})-\psi(n+\tfrac{d}{2})+2\psi(\tfrac{d}{2}+2n)\Big] \,.
\end{equation}



\end{document}